# Synchronization Processor Synthesis for Latency Insensitive Systems

Pierre Bomel, Eric Martin, Emmanuel Boutillon
*LESTER, Université de Bretagne Sud, Lorient, France*
*{pierre.bomel, emmanuel.boutillon, eric.martin}@univ-ubs.fr*

## Abstract

*In this paper we present our contribution in terms of synchronization processor for a SoC design methodology based on the theory of the latency insensitive systems (LIS) of Carloni et al[1]. Our contribution consists in IP encapsulation into a new wrapper model which speed and area are optimized and synthetizability guaranteed. The main benefit of our approach is to preserve the local IP performances when encapsulating them and reduce SoC silicon area.*

## 1. Introduction

Modern integrated circuits (ICs), named ''systems on a chip'' (SoCs), are the composition of several sub-systems exchanging data. SoC size increase is such that an efficient and reliable interconnection strategy is now necessary to combine sub-systems and preserve, at an acceptable design cost, the speed performances actual very deep sub-micron technologies provide[2]. This communication requirement can be satisfied by a LIS communication network between IPs. The LIS methodology enables to build functionally correct SoCs by 1) promoting pre-developed IPs intensive reuse, 2) segmenting inter-IPs interconnects with relay stations to break critical paths and 3) bringing robustness to data stream latencies to IPs by encapsulating them into synchronization wrappers. These encapsulated IPs are called "patient processes".

## 2. Related Works

Patient processes [3] are a key element in the LIS theory. They are suspendable synchronous IPs (named pearls) encapsulated into a wrapper (named shell) which function is to make them insensible to the IO latency and to drive the IP clock. The decision to drive or not the IP's clock is implemented very efficiently with combinatorial logic. Figure 1 illustrates a patient process' structure. Nevertheless, the LIS approach relies on a simplifying, but restricting, assumption: an IP is activated only if all its inputs are valid and all its outputs are able to store a result produced at next clock cycle. Now, it is frequent that only a subset of the inputs and outputs are necessary to execute one step of computation in a synchronous IP.

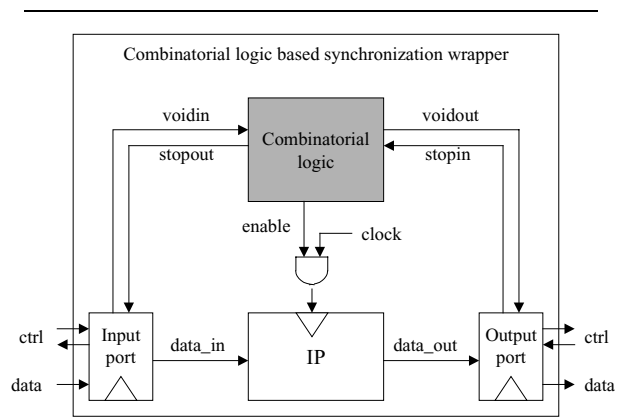

**Figure 1 : Carloni et al.'s Patient Process Model**

So as to limit the patient process sensitivity to a subset of the inputs and outputs, Singh and Theobald [4] suggest to replace the combinatorial logic that drives the clock by a Mealy type FSM. This FSM tests the state of only the relevant inputs and outputs at each cycle and drives the IP clock only when they are all ready. This approach is an extension of the LIS original model and has the advantage to correspond to a more realistic communication behavior. It can be implemented if one disposes of input/output schedules that proves the IP communication behavior is cyclic and not data-dependent : i.e. it is statically predictible. The major drawbacks of FSMs are their synthezability and silicon size when communication scenarios are long and complex like in computing intensive digital signal processing applications.

Finally, in order to reduce hardware cost, Casu and Macchiarulo [5] prove that, if it is possible to determine a static scheduling of all the IPs activation, then the relay stations can be replaced by simple flip-flops and the synchronization upstream and downstream protocol signals can be definitely removed. The IP activation static schedule is implemented with shift registers which contents drive the IP's clock. This approach relies on the hypothesis that there are no irregularities in the data streams: it is never necessary to randomly freeze the IPs.

## 3. New Approach – a Smaller Wrapper

As 1) LIS methodology lacks the ability to dynamically sense IO subsets, 2) FSMs can become too large as communication bandwidth does, and 3) shift register based synchronization targets only extremely rapid environments, we propose to encapsulate IPs into a new synchronization wrapper model which area is much less than the FSM-based wrappers area, speed is enhanced (mostly thanks to area reduction) and synthesizability is guaranteed whatever the communication schedule is.

The solution we suggest is functionally equivalent to the FSMs. This is a specific processor that reads and executes cyclically operations stored in a memory. We name it a "synchronization processor" (SP). Figure 2 specifies the new synchronization wrapper structure with our SP.

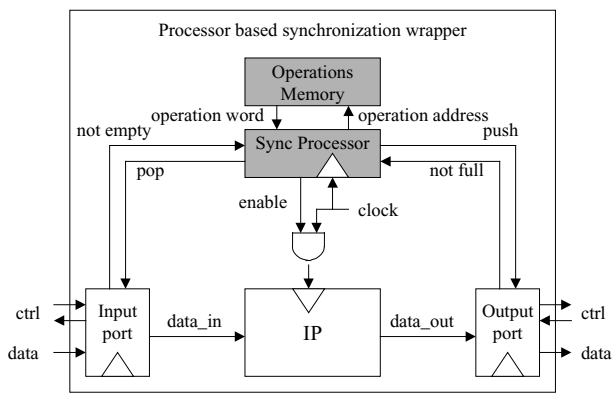

**Figure 2 : Patient Process Model with Synchronization Processor**

The SP communicates with the LIS ports with FIFO-like signals. These signals are formally equivalent to the *voidin/out* and *stopin/out* of [1] and *valid*, *ready* and *stall* of [4]. Number of input and output ports can be any. It drives the IP's clock with the *enable* signal. The SP model is specified by a three states FSM: a reset state at power up, an operation-read state, and a free-run state. This FSM is concurrent with the IP and contains a data path: this a "concurrent FSM with data path" (CFSMD). Operation's format is the concatenation of an input-mask, an output-mask and a free-run cycles number. The masks specify respectively the input and output ports the FSM is sensible to. The run cycles number represents the number of clock cycles the IP can execute until next synchronization point. To avoid unnecessary signals and save area, the memory is an asynchronous ROM (or SRAM with FPGAs) and its interface with the SP is reduced to two buses : the *operation address* and *operation word*. The execution of the program is driven by an operation "read-counter" incremented modulo the memory size.

## 4. Implementation and Results

Our SP has been successfully applied to the high-level synthesis of Reed-Solomon (RS) and Viterbi decoder IP cores with GAUT[6]. Table 1 gives comparative results of FSM and SP physical synthesis and shows that very important gains in area (up to 99 % saved) and speed (up to 47% increase) can be obtained with our SP.

| Complexity | FSM | | SP | | Gain (%) | |
|---|---|---|---|---|---|---|
| Port/wait/run | Sli. | Fr. | Sli. | Fr. | Sli. | Fr. |
| Viterbi 5 / 4 / 198 | 494 | 105 | 24 | 105 | -95 | 0 |
| RS 4 / 2957 / 1 | 2610 | 71 | 24 | 105 | -99 | +47 |

**Table 1 : Applicative Results**

## 5. Conclusion

Our SP has an essential characteristic: its complexity does not depend on the number of cycles the IP needs for a whole computation but only on the number of ports. Consequently its frequency and area are constant, for a given number of ports. It allows to build SoCs with a LIS based methodology, preserves IP frequencies and saves silicon space to implement patient processes.